# Exciton Delocalization Suppresses Polariton Scattering

Yongseok Hong[1], Ding Xu[1], Milan Delor[1,2,*]

[1]Department of Chemistry, Columbia University, New York, NY, 10027, US

[2]Lead contact: Milan E. Delor (milan.delor@columbia.edu)

**The bigger picture**
A central goal of chemistry and materials science is to develop avenues for efficient, coherent transport of electronic energy and information over long distances. Organic systems, such as molecular crystals and conjugated polymers, have shown promise in electronic devices but exhibit relatively poor transport due primarily to their localized electronic states. A frontier approach to enhance the transport of electronic excitations is to hybridize them with light, often achieved by embedding molecules or materials in resonant photonic structures. This strategy results in the formation of part-light part-matter particles known as polaritons, which inherit the long-range coherence of light with the ability to carry electronic energy and interface directly with electronic components. Nevertheless, despite this hybridization, molecular polaritons suffer from large disorder that leads to rapid scattering and dephasing, a major bottleneck to their implementation in polariton technologies. Here, we show that disorder-induced polariton scattering can be drastically suppressed in systems that possess some degree of intermolecular electronic coupling. We directly image polariton transport at light-like speeds using femtosecond optical microscopy, revealing that polaritons in molecular crystals and two-dimensional semiconductors exhibit exceptional transport over macroscopic scales even in the presence of large disorder. These results provide a general strategy and a predictive framework to optimize long-range polariton transport even in the presence of disorder, towards efficient polariton devices that simultaneously optimize sought-after coherence and strong interfacial interactions.

**Summary**
Exciton-polaritons (EPs) are part-light part-matter quasiparticles that combine large exciton-mediated nonlinearities with long-range coherence, ideal for energy harvesting and nonlinear optics. Optimizing EPs for these applications is predicated on a still-elusive understanding of how disorder affects their propagation and dephasing times. Here, using cutting-edge femtosecond spatiotemporal microscopy, we directly image EP propagation at light-like speeds in systems ranging from two-dimensional semiconductors to amorphous molecular films with systematically varied exciton-phonon coupling, exciton delocalization, and static disorder. Despite possessing similar EP dispersions, we observe dramatically different transport velocities and scattering times across systems. We establish a robust scaling law linking EP scattering to exciton transfer integral, demonstrating that polaritons based on materials with large exciton bandwidths are immune to disorder even for highly excitonic EPs. This observation cannot be deduced from the systems' linear optical properties, including EP dispersion and linewidth disorder. Our work highlights the critical and often-overlooked role of the matter component in dictating polariton properties, and provides precise guidelines for simultaneously optimizing EP propagation and nonlinearities.

**Introduction**

When light strongly couples to electronic transitions in molecular and material systems, new part-matter part-light eigenstates known as exciton-polaritons (EPs) can form.[1,2] EP formation is often facilitated by embedding excitonic systems in photonic cavities[3] that confine the light field and reduce photonic losses. EPs can inherit both the strong nonlinearity of excitons and the long-range coherence of light.[4,5] This hybrid nature has been leveraged to realize phenomena of high technological interest, including long-range electronic energy transport[6–15], polariton-assisted nonlinearities[16–21], low-threshold and high-temperature Bose-Einstein condensates[22,23], and single-photon blockades for quantum gates[24–26]. Optimizing these phenomena requires the simultaneous maximization of nonlinear interactions that occur at high excitonic content, and propagation or coherence lengths.

EP transport velocities are normally inferred from the gradient of the polariton dispersion. The picture typically invoked to describe EPs in the collective strong coupling regime is shown in Figure 1a, in which a cavity mode couples to $N$ excitons to form an upper polariton (UP), lower polariton (LP), and $N$-1 dark states (DS).[27,28] In this picture, the polariton dispersion (and velocity) is controlled by the cavity photon dispersion, Rabi splitting $\Omega_R$, and cavity-exciton detuning; the exciton dispersion contributes negligibly to the cavity polariton dispersion due to the small exciton bandwidth within the light cone. Nevertheless, recent studies have shown that the actual propagation velocities of polaritons can be much lower than their dispersions suggest, and that EPs exhibit transitions from ballistic (coherent) to diffusive (incoherent) transport at high excitonic fractions.[11,12,29–36] These results were assigned to disorder-induced polariton scattering or transient localization, either within the polariton manifold or into DS. These observations suggest a limited ability to simultaneously maintain long-range coherence and large nonlinearities that occur for highly excitonic EPs. A crucial question that remains unanswered is whether the properties of the underlying excitonic medium can be tuned to optimize this balance between propagation and nonlinear interactions that is central to EP applications.

Here, by directly imaging EP propagation in systematically varied materials using state-of-the-art ultrafast microscopy, we show that electronic delocalization in the underlying excitonic medium suppresses EP scattering even for high exciton fractions. We reveal a remarkably robust empirical scaling law linking the exciton transfer integral with EP scattering over a broad range of materials and temperatures. These results are not captured in prevalent models of polariton transport. We establish that materials possessing a rare combination of large exciton binding energies, oscillator strengths, and exciton transfer integrals, such as rigid two-dimensional (2D) perovskites, are ideal for simultaneously optimizing nonlinear interactions and long-range coherence at room temperature.

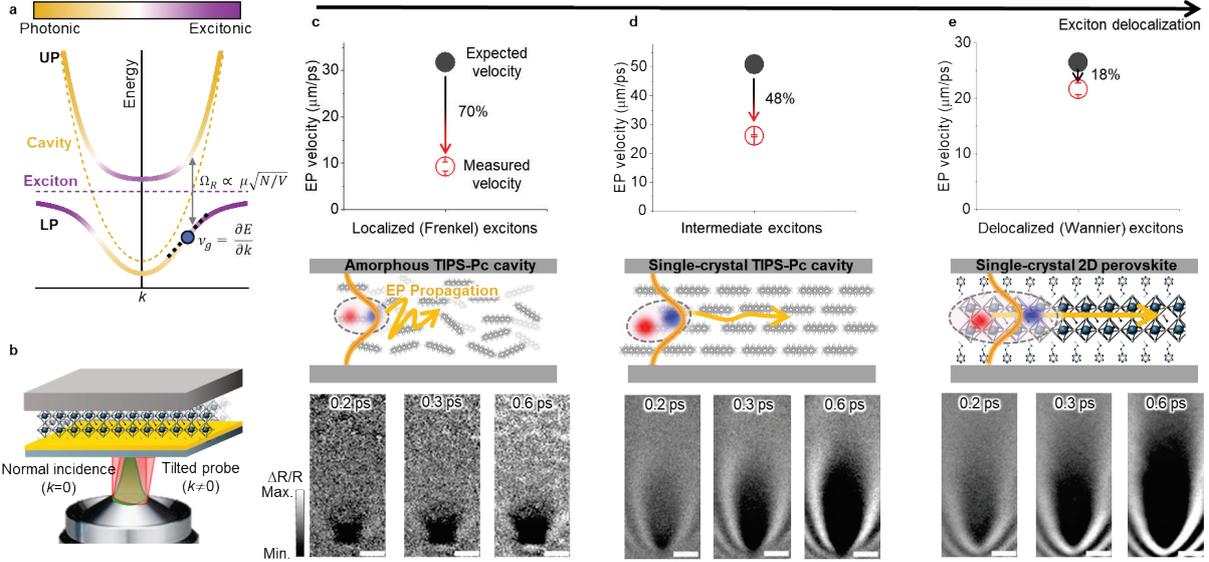

**Figure 1. Exciton delocalization protects EP transport. a.** Dispersion diagram for polaritons under strong light-matter coupling. The Rabi splitting $\Omega_R$ is related to the exciton transition dipole moment $\mu$, number of dipoles $N$ and cavity mode volume $V$. **b.** Schematic diagram for momentum-resolved ultrafast microscopy. **c-e.** EP transport dynamics imaged in a TIPS-pentacene microcavity (c, localized exciton), crystalline TIPS-pentacene microcavity (d, intermediate exciton), and crystalline 2D perovskite microcavity (e, delocalized exciton). The top panels display expected (grey filled circle) and measured (red circle) velocities for EPs with ~50% excitonic fractions in each cavity. The measured velocities are extracted from ultrafast optical imaging, with representative time delays shown on the bottom panel. TIPS-Pc stands for tri-isopropylsilyl-ethynyl pentacene. All data symbols are presented as the mean +/- one standard deviation derived from the fitting error. All scale bars, 1 μm.

## Delocalized excitons support delocalized polaritons

To characterize EP transport in different materials as a function of exciton fraction and temperature, we leverage momentum-resolved ultrafast polariton imaging (MUPI, Fig. 1b and Supplemental Note 1), a recently-developed spatiotemporally-resolved approach[11]. In MUPI, a diffraction-limited (~200 nm diameter) pump pulse generates a local nonequilibrium population of excitons and polaritons, and a widefield highly collimated monochromatic probe pulse impinges the sample at finite momentum $k$ to probe EPs at a specific point in their dispersion. The resulting pump-probe signal, collected in a reflection geometry, images the spatial evolution of polaritons with momentum-specificity through a nonlinear polariton blockade effect. In our current configuration, MUPI affords ~150 fs temporal resolution and < 30 nm spatial sensitivity to EP motion (Supplemental Note 1). This powerful combination of spatiotemporal, spectral and momentum specificity allows MUPI to be uniquely selective to EPs, avoiding complications from nonlinear signals associated with dielectric shift from exciton and nonequilibrium phonon populations[11].

Figures 1 c-e illustrate the transport properties for EPs with ~50% exciton fraction extracted from MUPI data in three microcavities. The microcavities comprise excitonic materials with progressively more delocalized excitons: an amorphous film of TIPS-pentacene molecules (Fig. 1c, localized excitons), a TIPS-pentacene molecular crystal (Fig. 1d, intermediate excitons), and the layered halide perovskite semiconductor $PEA_2(CH_3NH_3)Pb_2I_7$, where PEA is phenylethylammonium (Fig. 1e, delocalized excitons). We return to a detailed characterization of each system below. In all cases, we observe ultrafast EP flow over more than a micrometer within a picosecond. Nevertheless, Figures 1 c-e summarize a key finding of our study: despite each microcavity possessing similar polariton dispersions, we observe dramatic and visually salient differences in EP transport dynamics. Notably, the discrepancy between expected EP velocity (extracted from the experimentally-measured dispersion), and measured EP velocity (extracted from MUPI) depends monotonically on exciton delocalization. EPs in materials with more delocalized excitons result in propagation velocities closer to the expected group velocities. As detailed below, this trend cannot be explained by changes in EP dispersion, $\Omega_R$, cavity losses, or linewidth disorder. Instead, our results suggest that EP scattering and dephasing through interactions with phonons, disorder, or DS are suppressed in systems with delocalized excitons, despite the latter having no measurable impact on EP dispersion.

**Polarons renormalize polariton velocities**

Solution-processed halide perovskites have emerged as ideal systems to investigate polariton photophysics owing to their highly polarizable exciton transitions and strong nonlinearities.[19,37] To controllably evaluate the impact of exciton delocalization on EP scattering, we compare two halide perovskites with equivalent linear optical properties but distinct polaronic (non-perturbative exciton-phonon) interactions. We focus on the prototypical two-dimensional (2D) halide perovskites $A_2(CH_3NH_3)Pb_2I_7$ with A = PEA for a rigid lattice (PEAP) or butylammonium for a soft lattice (BAP), Figure 2a and Supplemental Note 2. PEAP and BAP possess near-identical exciton energies and linewidths (Fig. 2b, right inset), but the change in lattice rigidity induced by the different organic spacers leads to significant differences in excited-state exciton-phonon interactions. Using photoluminescence measurements (Supplemental Note 3), we estimate exciton–optical (OP) phonon coupling strengths of $\Gamma_{OP} = 25$ meV for PEAP in agreement with prior literature, compared to $\Gamma_{OP} = $ ~60 meV for BAP.[38] To realize strongly-coupled EPs, we embed mechanically-exfoliated single-crystal 2D perovskites in multimode Fabry-Pérot microcavities (Supplemental Note 2), as in our prior work that focused on BAP microcavities.[11] Angle-resolved reflectance spectra of the PEAP cavity dispersion in Figure 2b display clear LP branches, agreeing closely with transfer matrix simulations (Fig. 2b, right). Note that UP branches are not visible due to strong material absorbance above the semiconductor bandgap. Using a coupled oscillator model (dashed lines in Fig. 2b, and Supplemental Note 4), we extract a Rabi splitting of $\Omega_R = 260$ meV for PEAP, comparable to $\Omega = 275$ meV for BAP, with similar cavity quality (Q) factors of ~100 (Table S1).

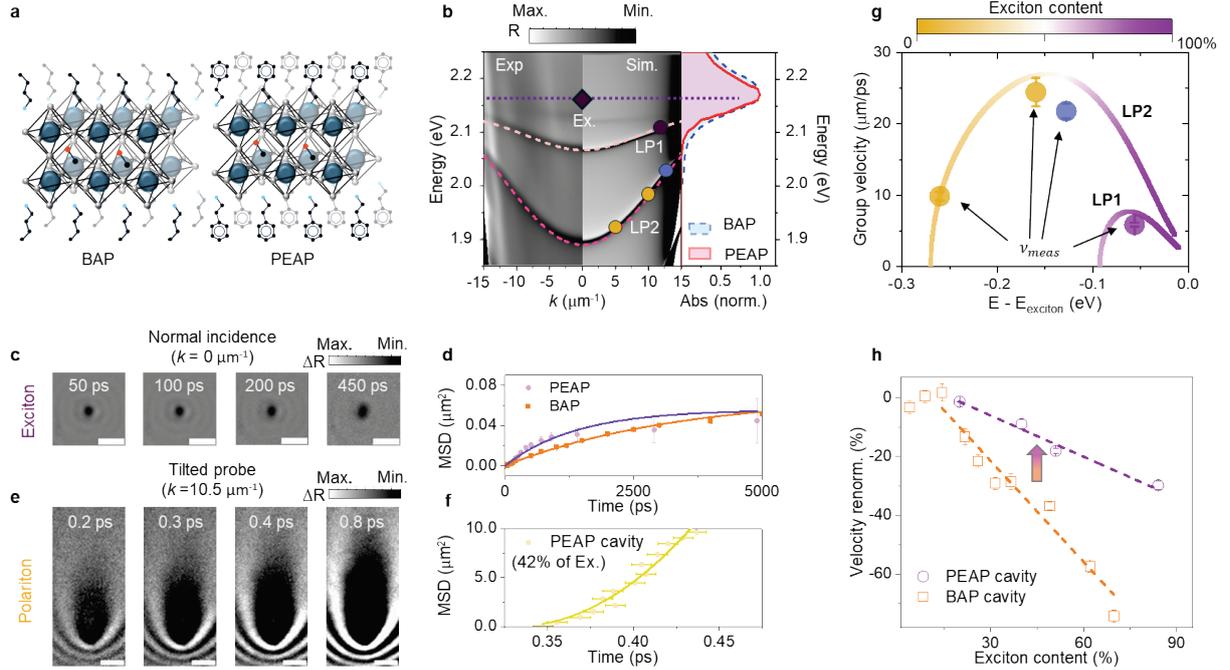

**Figure 2. EP transport in a 2D perovskite cavity. a.** Structures for BAP and PEAP 2D perovskites. **b.** Angle-resolved reflectance (left) and transfer-matrix simulations (right) of polariton dispersion for a PEAP microcavity. Dashed lines are coupled oscillator model fits. Right inset displays the linear absorbance for BAP and PEAP outside the cavity. **c.** Transport of reservoir excitons imaged at $k = 0$ μm$^{-1}$. **d.** MSD plot of exciton transport in PEAP and BAP microcavities. The result for BAP is from ref. 11. **e.** EP transport imaged at $k = 10.5$ μm$^{-1}$, corresponding to EPs with 42% excitonic fraction. **f.** MSD plot of EP transport from panel (e), extracted using wavefront analysis (Supplemental Note 5). **g.** Expected group velocity from the gradient of the polariton dispersion (solid line) vs. measured transport velocity (symbols) for the different points in the EP dispersion labeled with circles in Figure 2b. **h.** EP velocity renormalization as a function of exciton content for BAP (yellow square) and PEAP (purple circle) microcavities. The results for BAP are from ref.11. All pump-probe data is collected using focused 2.41 eV pump excitation with an incident peak fluence of ~5 μJ/cm$^2$. All data symbols are presented as the mean +/− one standard deviation derived from the fitting error. All scale bars, 1 μm.

Figure 2c displays spatiotemporal images tracking the propagation of uncoupled reservoir excitons in PEAP microcavities (probed at $k = 0$ μm$^{-1}$ filled diamond in Fig. 2b). From these images we extract the mean squared displacement MSD = σ$^2$(t) – σ$^2$(0), where σ$^2$(t) is the variance of the population spatial distribution at pump-probe delay $t$ extracted from a Gaussian fit to each image (Fig. 2d, analysis details in Fig. S6). For uncoupled excitons, we observe trap-limited transport characterized by an exponentially-decaying diffusivity, modeled as MSD = $2\lambda^2 \left[1 - \exp\left(-\frac{D}{\lambda^2}t\right)\right]$ (where D is diffusivity and $\lambda$ is the average distance between trap sites).[39] We extract $D = 0.21$ cm$^2$/s and a 2D trap density (1/$\lambda^2$) of 35.7 μm$^{-2}$. Figure 2d also plots the MSD for uncoupled excitons in BAP microcavities, which show similarly trap-limited transport, but with lower diffusivity $D = 0.10$ cm$^2$/s and trap density of 29 μm$^{-2}$.[11] These values concur with literature values of exciton transport in BAP and PEAP crystals outside microcavities.[39–41] The larger exciton diffusivity in PEAP compared to BAP is attributed to lower exciton-phonon coupling and a more rigid lattice in the former.[39]

Figure 2e shows MUPI images of EP transport in the lower polariton branch for the PEAP microcavity probed at $k = 10.5$ μm$^{-1}$ (yellow circle in Fig. 2b), corresponding to an excitonic fraction of 42%. All MUPI data presented herein is collected using a non-resonant (above-gap) focused pump; we previously showed that EP velocities are equivalent under resonant and non-resonant excitation conditions[11]. The MSD for EP propagation (Fig. 2f) is quadratic in time, indicating ballistic (scatter-free) transport with a group velocity of $v_{\text{meas}}=$ 24 μm/ps, or ~8% of light speed. These measurements confirm that EPs realized through cavity strong coupling enable orders-of-magnitude enhancement in electronic energy transport, as previously observed.[9,11,12,29,42–44]

In Figure 2g, we plot the measured EP velocity (symbols) as a function of exciton content by leveraging the momentum-selectivity of MUPI to probe different points in the polariton dispersion. The measured EP velocities are overlaid with the expected group velocity (Fig. 2g, colored lines) extracted from the gradient of the experimentally-measured dispersion in Fig. 2b. The dependence of EP transport on exciton content is in agreement with previous work on BAP[11] and organic microcavities[12] – we observe: (i) An increasingly large deviation between $v_{\text{meas}}$ and the expected group velocity ($v_{\text{g}}$) as the exciton content increases due to polariton scattering; (ii) A ballistic-to-diffusive transition for EPs with exciton content higher than 50% (Fig. S8).

The key comparison of EP velocities between PEAP and BAP cavities is shown in Figure 2h, which plots the velocity renormalization (defined as $v_{\text{renorm.}} = 100 \times (v_{\text{meas.}} - v_{\text{g}})/v_{\text{g}}$), as a function of exciton content. Crucially, we observe a clear difference in the degree of renormalization between the two systems, with BAP experiencing ~2.8 fold greater renormalization compared to PEAP. We emphasize that the factors typically assumed to affect EP transport (Rabi splitting, cavity Q-factor, and linewidth of the exciton transition) are within experimental error of each other for the two materials (Table S1), indicating that these features are not sufficient to capture the different EP velocity renormalizations between PEAP and BAP cavities. We recover EP velocities that are renormalized by only ~30% for EPs with 87% exciton content in PEAP, indicating that velocity renormalization can be minimal even at room temperature and in standard Fabry-Perot microcavities with an appropriate material choice.

Our results indicate that EP transport depends heavily on underlying exciton-phonon interactions. The fact that these interactions are not captured in the linear polariton dispersion or optical spectra suggest that they are non-Condon excited-state interactions. Specifically, 2D halide perovskites are known to undergo exciton-polaron formation[45–47], wherein the lattice relaxes around the exciton to form an electronic quasiparticle dressed by a lattice deformation, resulting in an increase of the exciton effective mass and reduced exciton delocalization.

Although polaron formation is likely to occur in both BAP and PEAP crystals, the lattice relaxation is greater in the more flexible BAP (Supplemental Note 7).[39] We therefore assign the different EP transport properties for BAP and PEAP observed in Figure 2h to EP–lattice interactions mediated by polaron formation. These results confirm – using a well-controlled material platform – that the details of the underlying excitonic medium strongly affects EP properties, beyond the combination of dispersive photons and non-dispersive excitations captured in static optical properties.

**Exciton-phonon and polariton-phonon interactions are different**

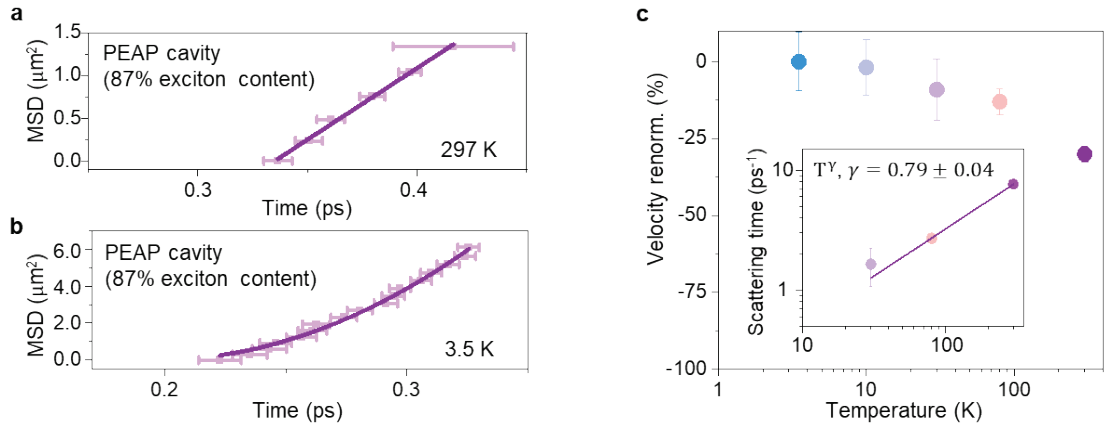

**Figure 3. EP-phonon interactions in 2D perovskite microcavities**. **a, b**. MSD plot for EPs with 87% exciton fraction in PEAP microcavity at 297 K (a) and 3.5 K (b). **c**. Temperature-dependent velocity renormalization for PEAP microcavity with 87% exciton content. The inset is the scattering rate plotted as a function of temperature. All data symbols are presented as the mean +/− one standard deviation derived from the fitting error.

Although results presented above indicate an intimate connection between exciton-lattice and EP-lattice interactions, the question of whether the same phonons are involved in both remains an open question. To answer this question, we perform temperature-dependent measurements of EP transport in PEAP cavities. First, we find that even highly excitonic EPs that display diffusive transport at room temperature (Fig. 3a for 87% exciton content) recover ballistic transport at 3.5 K (Fig. 3b) with no velocity renormalization. This result reinforces our assignment of velocity renormalization to EP-phonon interactions, and indicate that these interactions give rise to decoherence at elevated temperatures for highly excitonic EPs.

The temperature-dependent renormalization plotted in Figure 3c provides further insight on EP-lattice interactions. To extract an effective EP-phonon scattering time, we assume a semiclassical picture $2D = v_g^2 \times \tau_{\text{scattering}}$ and extract $D$ assuming linear MSDs at each temperature. In this picture, the scattering rate normally obeys power law behavior ($1/\tau_{\text{scattering}} \propto T^\gamma$), where the value of the exponent characterizes the dominant phonon contribution.[48,49] For example, in 2D, acoustic phonon scattering is characterized by $\gamma = 1$.[49]

Outside the microcavity, PEAP exhibits characteristic values of $\gamma$ between 2.1 – 2.6 (Fig. S9), indicating that scattering is primarily mediated by out-of-plane non-polar optical phonons ($\gamma > 1$), in line with prior results.[48,50] In contrast, under strong light-matter coupling in microcavities, we observe $\gamma = 0.79 \pm 0.04$ (inset in Fig. 3c) over 30 - 300 K. This value of $\gamma$ indicates a substantially reduced contribution from short-range optical phonons in favor of longer-range acoustic phonon scattering. This observation suggests that the delocalization of EPs results in lattice interactions that can differ from those of bare excitons, as was recently theoretically postulated[36]. This result reinforces a picture of EPs as hybrid quasiparticles that go beyond linear sums of their underlying components, and calls for models of EP transport that go beyond projecting known exciton-phonon interactions into the polariton basis.

**Robust scaling law links polariton scattering to exciton transfer integral**

A largely overlooked parameter that can affect EP transport is the extent of exciton delocalization, as shown in Figure 1. Here, we show that the exciton transfer integral $J$ is likely the best predictor for the degree of EP velocity renormalization experienced by all systems investigated in this work. Identifying systems with radically different $J$ to compare EP transport is challenging because other properties such as linewidth inevitably change. We identify tri-isopropylsilyl-ethynyl pentacene (TIPS-Pc) as a promising candidate because: (i) We can prepare microcavities from large single-crystal domains with $J \approx 43.9$ meV (Figs. 4a and S12, see Supplemental Note 7 for estimation of $J$), and positionally-disordered monomeric molecules in a polymer matrix ($J \approx 0$ meV, Fig. 4b) with comparable Rabi splitting. (ii) The slight increase in exciton linewidth in the single-crystal microcavities (~128.2 meV compared to ~100.9 meV in the monomer film) allows us to rule out that larger EP renormalization in the low-$J$ sample is due to larger linewidth $\Gamma$, or a lower ratio $\Omega_R/\Gamma$, which is currently understood as a good predictor of EP delocalization.[51–53]

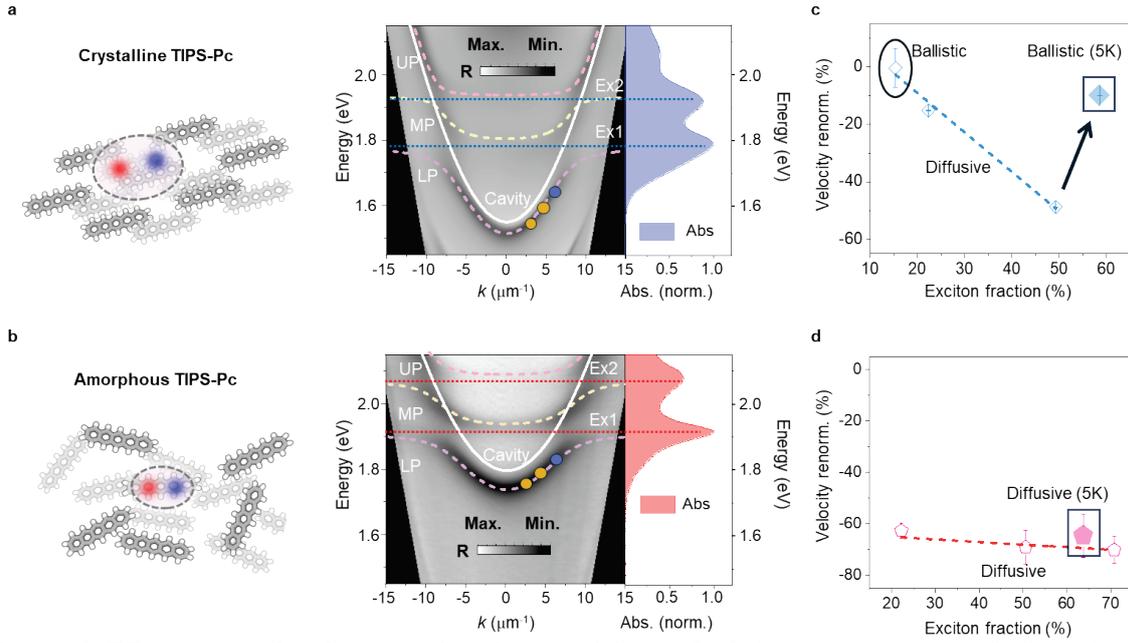

**Figure 4**. **EP propagation in organic microcavities. a, b.** Schematic structure (left) and angle-resolved reflectance spectrum showing the polariton dispersion (right) for a crystalline TIPS-Pc microcavity (a) and amorphous TIPS-Pc film (b). Right insets display the linear absorbance for each sample outside the cavity. **c.** EP velocity renormalization as a function of exciton content in the crystalline TIPS-Pc microcavity (blue diamond). The filled diamond indicates the measured EP velocity at 5 K. **d.** EP velocity renormalization as a function of exciton content in the amorphous TIPS-Pc microcavity (red pentagon). The filled pentagon indicates the measured EP velocity at 5 K. All pump-probe data is collected using focused 1.97 eV pump excitation with an incident peak fluence of ~2 μJ/cm$^2$. All data symbols are presented as the mean +/− one standard deviation derived from the fitting error.

As shown in Figures 4a,b, distinct upper (UP), middle (MP), and lower (LP) polaritons are observed in both single-crystal and amorphous molecular films. From coupled oscillator model fits, we extract $\Omega_R$ = 200 meV and 160 meV for TIPS-Pc crystalline and amorphous cavities, respectively. Although the cavity–exciton detunings for the two cavities investigated here are different, we have previously demonstrated that the effect of cavity detuning on EP renormalization for a given excitonic fraction is negligible[11]; note that the exciton fraction determines the actual detuning between the EP being probed and the exciton reservoir regardless of cavity detuning. Importantly, both cavities show similar ratios $\Omega_R/\Gamma$ = 1.59 for the crystalline system and 1.56 for the amorphous system.

Despite these similarities, we observe drastically different EP transport for crystalline vs amorphous samples both at room temperature and at 5 K. The EP velocity renormalization in the TIPS-Pc crystal cavity follows a similar trend to the perovskite cavities, but with a 3.2-fold steeper dependence on exciton content compared to PEAP, reaching over 50% renormalization for 49% exciton content (Figs. 4c and S14). In contrast, the TIPS-Pc amorphous cavity exhibits a large, relatively uniform velocity renormalization of 60-70% across exciton contents spanning 22-71% (Figs. 4d and S15). At 5 K, further differences emerge: the crystalline cavity recovers ballistic transport with a non-zero but small velocity renormalization (Figs. 4c and S16), while the amorphous cavity maintains diffusive transport with renormalization comparable to room temperature data (Figs. 4d and S17). The crossover from diffusive to ballistic transport as the temperature lowers indicates that dynamic disorder dominates EP scattering in the crystalline system, similarly to the perovskite cavities. In contrast, the large velocity renormalization and persistent diffusive transport at 5 K in the amorphous cavity signal the inability of EPs to overcome scattering from static disorder. We emphasize that changes in the linear absorption properties of amorphous and crystalline samples as a function of temperature are unable to explain these striking differences (Fig. S18).

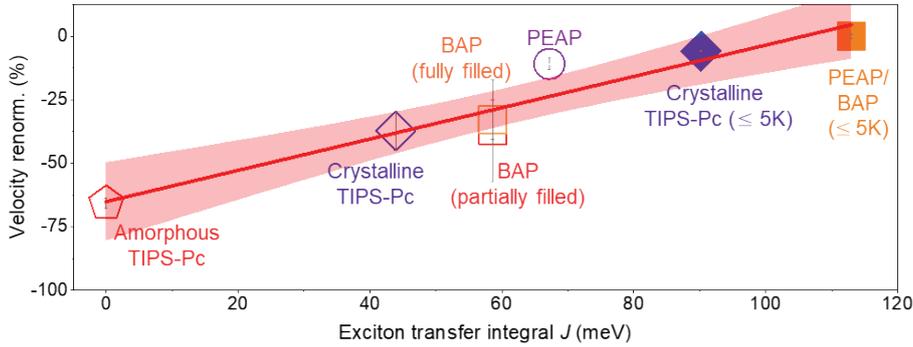

**Figure 5. EP scattering depends on excitonic intersite coupling.** Summary of velocity renormalization for EPs with 40% exciton fraction plotted as a function of intersite exciton transfer integral for PEAP (purple circles), BAP (yellow squares), crystalline TIPS-Pc (blue diamonds), and amorphous TIPS-Pc (red pentagons) microcavities. Open symbols indicate room temperature data; filled symbols indicate low-temperature data. Estimations of $J$ for different temperatures are discussed in Supplemental Note 7. The shaded region indicates the 95% confidence band of a linear fit to the data. The plot displays a clear correlation between intersite excitonic coupling and EP transport.

Overall, our results indicate that polariton dispersions are insufficient to predict EP transport properties in a large range of systems due to EP scattering, transient localization, and decoherence on much shorter scales than the polariton lifetime for highly excitonic EPs. Nevertheless, a clear trend emerges from our measurements: higher electronic coupling in the underlying excitonic medium leads to better EP transport, which we define as lower EP renormalization for a given exciton content. To capture this trend quantitatively, Figure 5 plots the velocity renormalization for EPs with 40% exciton character as a function of $J$, for all measured systems at room temperature (open symbols) and low temperature (solid symbols). At low temperature, $J$ is larger in TIPS-Pc, BAP and PEAP crystals, which all experience renormalized masses at room temperature due to polaron formation (Supplemental Note 7). Figure 5 shows that the relationship between EP renormalization and $J$ is maintained over all data collected in this work, cementing our findings as a robust empirical scaling law. We find much poorer correlations between EP renormalization and other common parameters such as Q-factor, $\Omega_R$ or $\Gamma$, including when comparing BAP microcavities possessing different $\Omega_R$ tuned through cavity geometry (Supplemental Note 16).[54] Therefore, the key finding from this plot is that EP scattering for highly excitonic EPs can be dramatically suppressed in systems that possess delocalized excitons. Remarkably, within the systems explored here, the strongest predictor for whether EP transport reaches its expected group velocity is the underlying excitonic band structure.

One way to rationalize the observed scaling between EP renormalization and $J$ is to invoke a picture where phonons and/or static disorder mediate EP scattering into relatively localized states such as DS, including through a recently-proposed superexchange mechanism that does not require transferring population to DS[55]. In this picture, large $J$ implies more delocalized dipoles, which in turn reduces the density of DS under collective light-matter coupling and associated EP-DS scattering. Similarly, a recently-developed theoretical framework suggests that large intersite exciton coupling in 2D molecular aggregates protects polaritons against disorder through persistent DS delocalization.[56] Taken together, our observations indicate that delocalized excitons in the underlying medium allows highly excitonic EPs to maintain long-range coherence and transport. Achieving sufficient oscillator strengths in delocalized (non-Frenkel) excitonic systems to allow strong coupling at room temperature can be challenging. Layered semiconductors such as 2D halide perovskites are ideal in this respect, as they combine macroscopic crystallinity with large oscillator strengths and appreciable exciton delocalization, with the latter optimized in structures that are more rigid.

Through systematic imaging of EP transport in a variety of microcavity polaritons, we have elucidated to what extent the details of the underlying excitonic medium project into the properties of hybridized EPs. Our most important finding is that exciton-lattice interactions and excitonic intersite coupling are crucial predictors of EP transport, even when these features have no influence on the linear optical and dispersion properties of EPs. Most importantly, we reveal a robust empirical scaling law linking EP velocity renormalization to exciton delocalization. This relationship spans from disordered organic molecular films to highly crystalline 2D semiconductors over a large temperature range, and should extend to new polaritonic material platforms[57,58]. Our results underscore the importance of the underlying matter in dictating EP properties, and provides guidelines for designing excitonic systems that maximize both nonlinear interactions and long-range coherence, a key demand of future polaritonic devices.

## Resource availability

### Lead contact
Further information and requests for resources should be directed to the lead contact, Milan Delor (md3864@columbia.edu).

### Data availability
All raw data are displayed in Figs. 1–5 of the main text and Figs. S1–S20 of the Supplemental Information. Raw image files are available from the contact author upon request.

### Acknowledgements
This material is based upon work primarily supported by the National Science Foundation under Grant Number CHE-2203844, and by the Arnold and Mabel Beckman Foundation through a Beckman Young Investigator award. Low-temperature measurements were supported by the National Science Foundation under Grant Number DMR-2115625. Y.H. acknowledges support from the National Research Foundation of Korea (NRF) grant funded by the Korea government (MSIT) (RS-2023-00240362).

### Author contributions
Y.H. and M.D. conceived and designed the experiments. Y.H. and D.X. prepared samples. Y.H. acquired the experimental data. Y.H. and D.X. analysed the experimental results and performed the numerical simulations. M.D. supervised the project. Y.H. and M.D. wrote the paper with input from all authors.

### Declaration of interests
The authors declare no competing interests.

Supplemental Information for:

Exciton Delocalization Suppresses Polariton Scattering

Yongseok Hong[1], Ding Xu[1], and Milan Delor[1,2]*

[1.] Department of Chemistry, Columbia University, New York, NY 10027, United States
[2.] Lead Contact: Milan Delor (milan.delor@columbia.edu)

**Supplemental text**

## Note 1. Ultrafast momentum-resolved polariton imaging (MUPI)

Exciton polariton imaging of various Fabry-Pérot microcavities is performed using ultrafast momentum-resolved polariton imaging (MUPI). The systems used here are described in detail elsewhere[1,2] and illustrated schematically in Figure S1.

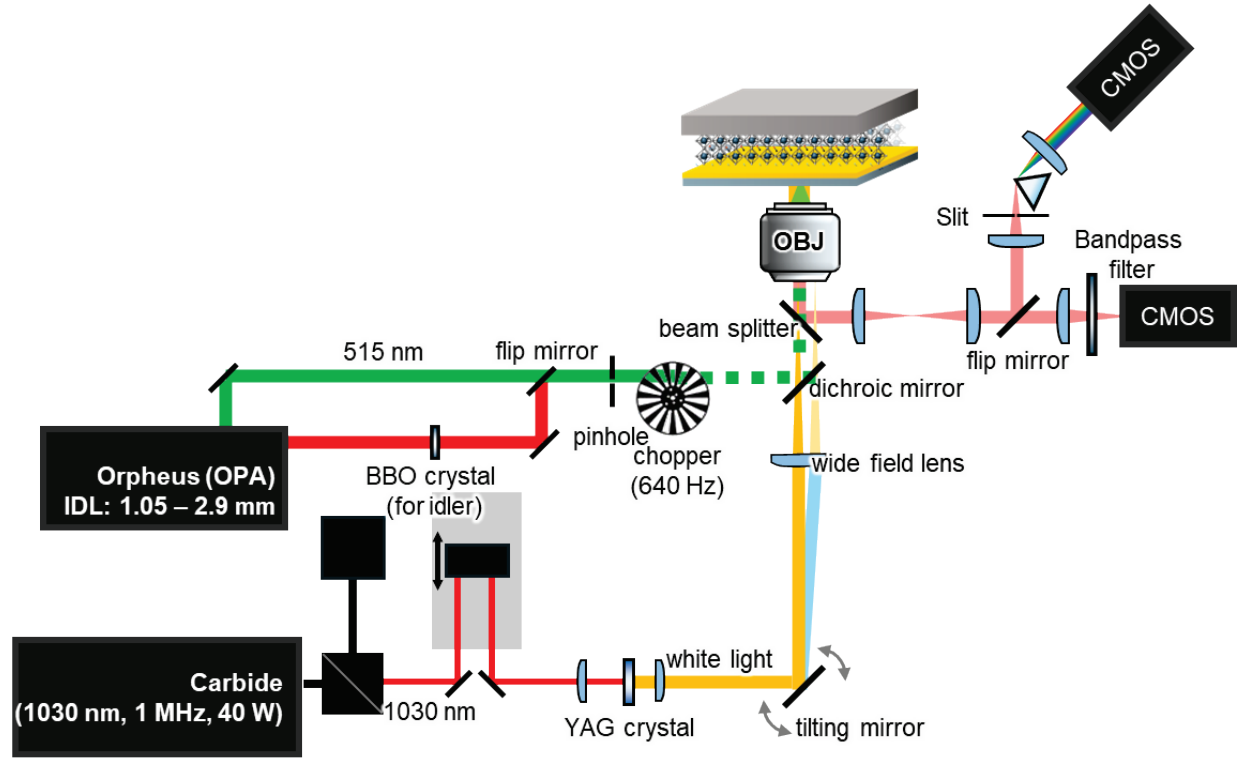

**Figure S1.** MUPI setup schematic.

Briefly, a Yb:KGW ultrafast regenerative amplifier (Light Conversion Carbide, 40 W, 1030 nm fundamental, 1 MHz repetition rate) was used as the light source for spatiotemporal measurements. The pump pulses are generated by frequency-doubling either the fundamental 1030 nm output, or the idler output of an optical parametric amplifier (OPA, Light Conversion, Orpheus-F). A white light continuum (WLC) as a probe pulse is generated in a YAG window (EKSMA). The pump and probe beams are directed to the high numerical-aperture oil-immersion objective (Leica HC Plan Apo 63x, 1.4 NA oil immersion). The pump beam is collimated before the objective lens and focuses on the sample plane. The probe beam focuses on the back focal plane of the objective by an $f = 300$ mm widefield lens, enabling widefield illumination of the sample. A tilting mirror placed one focal length upstream of the widefield lens allows tuning the angle at which the widefield probe illuminates the sample, thus allowing probing at any momentum up to a maximum of $k/k_0 = 1.4$, limited by the numerical aperture of the objective. The reflected probe light at the sample interface with the back-scattered light from the sample forms images on a CMOS camera (Blackfly S USB3, BFS-U3-28S5M-C). For cryogenic measurements, the same light source is used, but the sample is placed in a closed loop Montana Instruments s100 cryostation equipped with a cryo-optic objective (Zeiss LD EC Epiplan-Neofluar 100x/0.90 DIC M27, NA = 0.9). For angle-resolved reflectance, the back focal plane of the objective is projected on the entrance slit of a home-built prism spectrometer using a pair of lenses ($f_1 = 300$ mm and $f_2 = 100$ mm), as depicted in Figure S4. For real-space MUPI imaging, an additional 150 mm lens forms an image on a CMOS camera (Blackfly S USB3, BFS-U3-28S5M-C). Both the spectrometer camera and the real-space camera are triggered at double the pump modulation rate, allowing the consecutive acquisition of images with the pump ON followed by the pump OFF. Consecutive frames are then processed according to (pump on/pump off – 1). The pump-probe temporal cross-correlation at the sample is approximately 166 fs (Figure S2).

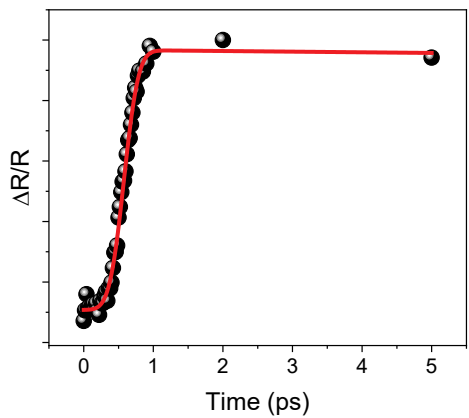

**Figure S2**. Normalized pump-induced reflection profile using 515 pump and 620 nm probe. The estimated instrument response function is 166 fs.

Here, we note that the spatial precision for both exciton and polariton transport measurements is not defined by the diffraction limit, but rather by the signal-to-noise ratio (SNR) of the measurement, as is well described in the field of spatiotemporal imaging.[3,4] Our measurements typically yield sub-30 nm spatial precision.

**Note 2.      Sample preparation**

**$(C_6H_5C_2H_7N)_2(CH_3NH_3)Pb_2I_7$ ($PEA_2(MA)Pb_2I_7$) crystal synthesis.** The synthesis of large-area single-crystal flakes followed a published procedure.[1,5] All chemicals were purchased from Sigma-Aldrich. Hydriodic acid (HI) solution was prepared by mixing 57% wt. aqueous HI (9 mL) and 50% wt. aqueous Hypophosphorous acid ($H_3PO_2$, 1 mL). Lead (II) iodide ($PbI_2$, 99.999% trace metals bases) powder (2720 mg, 5.9 mmol) was dissolved in the solution mixture under constant magnetic stirring. Methylammonium iodide (MAI, ≥ 99%, anhydrous 493 mg, 3.1 mmol) was added to the solution, and black powder precipitated instantly. The black powder redissolved quickly by heating the solution to 100 °C with stirring. After subsequent addition of phenylethylammonium iodide (PEAI, 864 mg, 4.3 mmol), the solution was heated to 105 °C under constant magnetic stirring until all precipitate dissolved. The solution was then subjected to a controlled cooling rate of 0.5 °C/h to room temperature in an oil bath, and large crystals formed on the solution surface. The crystals were collected by vacuum filtering and washed twice with toluene.

**Metallic cavity fabrication.** The bottom partial reflector through which light impinges the sample is a 30 nm gold film that was deposited on cover glass (Richard-Allan Scientific, 24×50 #1.5) by e-beam evaporation (Angstrom EvoVac deposition system). The deposition rate was 0.05 nm/s. The perovskite flakes were then mechanically exfoliated onto the gold-deposited cover glass with PVC tape (Nitto SPV224 PVC Vinyl Surface Protection Specialty Tape). Although gold is more lossy than silver, thin (~50nm) silver films tend to deteriorate during measurements, which rapidly leads to poor imaging quality, an important aspect for our scattering based experiments. The top reflector is a 200 nm silver film that was prepared by e-beam deposition on a silicon wafer. A layer of polyvinylpyrrolidone (PVP) solution (Sigma Aldrich, M.W. 40000, 10% wt in ethanol/acetonitrile wt 1/1) was then spin-coated on the silver film (3000 rpm, acceleration 1000rpm/s, 2min) and thermally annealed at 150 °C for 5 min. The prepared PVP/Ag was picked up with thermal release tape (semiconductor corp., release temperature 90 °C) and placed on the perovskites flakes with firm pressure, completing the cavity structure. The full structures were encapsulated between glass slides using epoxy (OG159-2, Epoxy Technology) in a nitrogen-filled glovebox to prevent sample oxidation during the measurements.

**Amorphous/Crystalline TIPS-pentacene cavities.** For amorphous films, TIPS-Pentacene (≥ 99%, 20 mg) and polystyrene (PS, M.W. 280000, 60 mg) were dissolved in toluene (1 mL) under constant stirring at 60ºC until the TIPS-pentacene/PS is fully dissolved. The TIPS-pentacene/PS solution is then spin-coated on the gold coated cover glass at 3000 rpm for 120 seconds. In contrast, for a crystalline film, TIPS-Pentacene (5 mg) was dissolved in toluene under constant stirring at 60ºC. 50 ul of TIPS-pentacene solution was then drop-cast onto the gold coated cover glass. TIPS-pentacene crystals with large domain sizes form within 20 minutes of drop-casting. For both amorphous and crystalline systems, we then deposit a 200 nm gold film by e-beam evaporation on the samples. The deposition rate was 0.1 nm/s.

## Note 3. Temperature dependent photoluminescence linewidths of PEAP

We perform temperature-dependent photoluminescence (PL) experiments on PEAP single crystals (Figure S3a). We analyze the linewidth evolution as a function of temperature to estimate the exciton-phonon coupling strength of PEAP crystals (Figure S3b). Exciton-phonon interactions are captured in the Debye-Einstein approximation:[6,7]

$$\Gamma = \Gamma_0 + \Upsilon_A T + \frac{\Gamma_{OP}}{\exp\left(\frac{E_0}{k_B T}\right) - 1} \tag{1}$$

where $\Gamma_0$ is PL line width at 0 K, $\Upsilon_A T$ is the broadening induced by acoustic phonons, and $\Gamma_{OP}$ is the interaction with optical phonons. By fitting the data in Figure S3b, we extract coupling strengths for acoustic and optical phonons of 87 µeV/K and 24.6 meV, respectively, in line with previous results.[6,8]

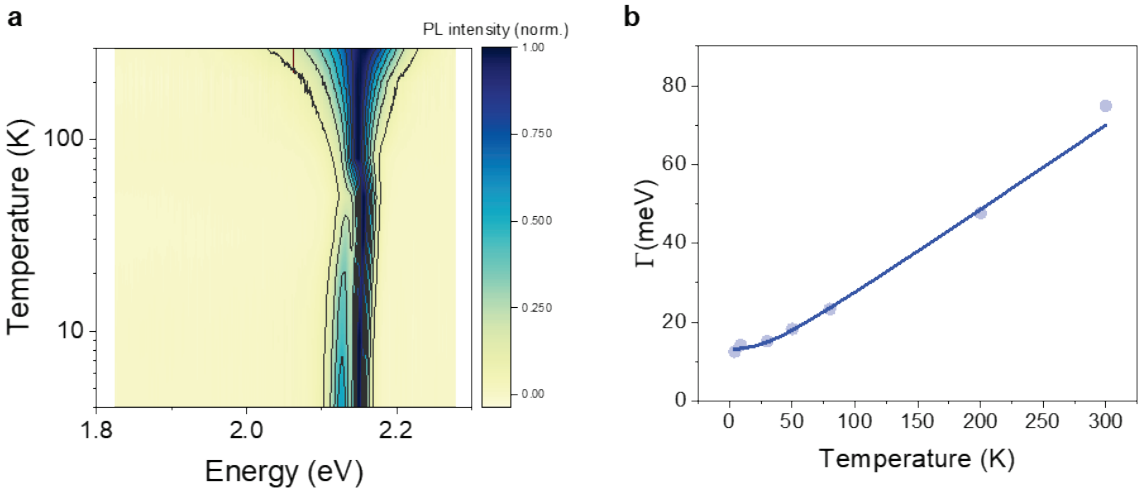

**Figure S3. Temperature dependent PL results for PEAP. a.** Contour map of normalized PL as a function of temperature. **b.** Full-width half maximum (FWHM) of PL spectrum as a function of temperature. The solid line represents the Debye-Einstein fit to extract contributions from phonon-induced broadening.

# Note 4. Analysis of polariton dispersions

Here, we detail the coupled oscillator models used to model cavity dispersion for various geometries[9–12] (Figure S4).

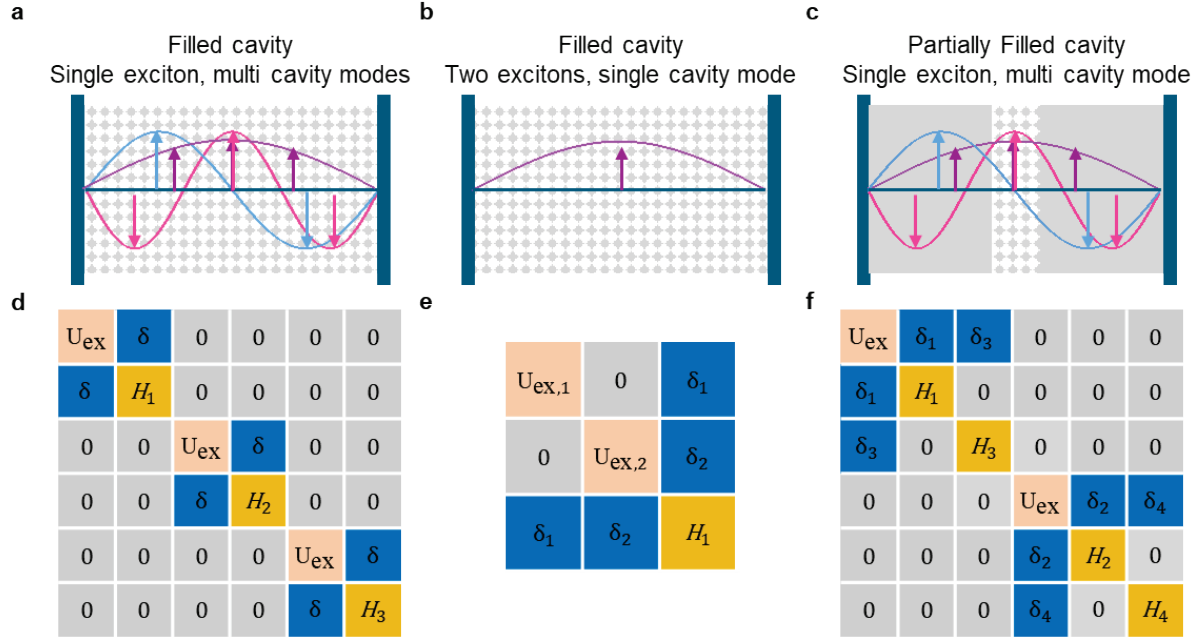

**Figure S4.** Schematic illustrations of various cavities (**a-c**) and their corresponding light–matter matrices (**d–f**) that can be diagonalized to obtain multimode polariton dispersion. The dotted box corresponds to the materials with various thicknesses and the solid box indicates the layer of PMMA polymer.

Fully-filled multi-mode perovskite cavities in the strong-coupling regime (Figure S4a) are best fit using a coupled oscillator model described by a $2N$-dimension block-diagonal Hamiltonian for $N$ cavity modes. For the three modes observed in the dispersion in our case, we model the dispersion using the following coupled oscillator model:

$$\begin{bmatrix} U_{ex} & \delta & 0 & 0 & 0 & 0 \\ \delta & H_1 & 0 & 0 & 0 & 0 \\ 0 & 0 & U_{ex} & \delta & 0 & 0 \\ 0 & 0 & \delta & H_2 & 0 & 0 \\ 0 & 0 & 0 & 0 & U_{ex} & \delta \\ 0 & 0 & 0 & 0 & \delta & H_3 \end{bmatrix} \begin{bmatrix} \chi_1 \\ \varphi_1 \\ \chi_2 \\ \varphi_2 \\ \chi_3 \\ \varphi_3 \end{bmatrix} = E_{pol} \begin{bmatrix} \chi_1 \\ \varphi_1 \\ \chi_2 \\ \varphi_2 \\ \chi_3 \\ \varphi_3 \end{bmatrix} \quad (2)$$

where $U_{ex}$ is the exciton energy (2.14 eV), $H_n$ is the energy for each cavity mode, $\delta$ is the exciton-photon coupling ($\Omega_R/2$) between each cavity mode and the exciton, and $E_{pol}$ is the energy of polariton. Figure 2b (dashed lines) and Figure S5 display the result of such a coupled oscillator fit for PEAP cavities. The fit provides a value of $\delta$ = 130 meV, i.e. a Rabi splitting of 260 meV.

For TIPS-Pc single-mode cavities (Figure S4b), in which the excitonic absorption shows distinct vibronic features, the lowest two vibronic bands both couple with the cavity modes. These cavities are described using:

$$\begin{bmatrix} U_{ex,1} & 0 & \delta_1 \\ 0 & U_{ex,2} & \delta_2 \\ \delta_1 & \delta_2 & H_1 \end{bmatrix} \quad (3)$$

where $U_{ex1\ and\ 2}$ correspond to the lowest two vibronic energies (1.78 and 1.9 eV for crystalline TIPS-Pc and 1.9 and 2.05 eV for amorphous TIPS-Pc). The fits (Figures 4a and 4b) provide values of $\delta_1$ = 80 (100) meV for amorphous (crystalline) TIPS-Pc, i.e. a Rabi splitting of 160 and 200 meV, respectively.

Finally, for the partially-filled multimode BAP cavity (PMMA/perovskite/PMMA, Figure S4c), it is necessary to distinguish the light-matter coupling with even- and odd- cavity modes. This type of cavity is best modeled using an $(N+2)(N+2)$-dimension Hamiltonian for $N$ cavity modes:

$$\begin{bmatrix} U_{ex,1} & \delta_1 & \delta_3 & 0 & 0 & 0 \\ \delta_1 & H_1 & 0 & 0 & 0 & 0 \\ \delta_3 & 0 & H_3 & 0 & 0 & 0 \\ 0 & 0 & 0 & U_{ex,1} & \delta_2 & \delta_4 \\ 0 & 0 & 0 & \delta_2 & H_2 & 0 \\ 0 & 0 & 0 & \delta_4 & 0 & H_4 \end{bmatrix} \quad (4)$$

The fit (Figure S20) provides a value of $\delta$ = 90 meV, i.e. a Rabi splitting of 180 meV.

Figure S3b shows the result of a scattering matrix method (SMM) calculation, performed using the open-source S4 package,[13] using the perovskite slab thickness and the known dielectric functions of the metallic mirrors and the perovskite. The experimental dispersion, coupled oscillator models (Figure S5) and SMM calculations are all in excellent agreement with one another. Note that upper polariton branches are not visible in the experimental dispersion (Figure S5a) because the material is highly absorptive above bandgap, as also captured in the SMM calculation (Figure S5b).

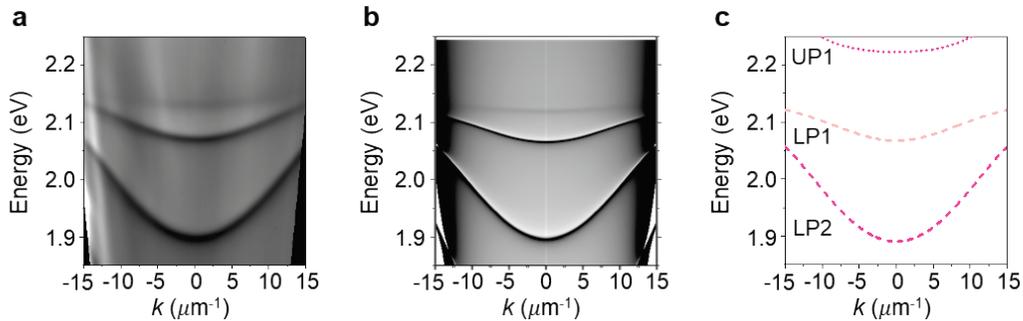

**Figure S5.** **a.** Angle-resolved reflectance of dispersion spectrum for PEAP microcavity. **b.** SMM calculation. **c.** Coupled oscillator model fits.

The exciton content of the polariton for each mode (Hopfield coefficients) are obtained from eigenstate normalization.

The exciton transition linewidth, cavity-exciton detuning, Rabi splitting, and Q factors are tabulated in Table S1.

**Table S1. Linewidth of exciton transition ($\Gamma$), cavity-exciton detuning ($\Delta E$), Rabi splitting ($\Omega_R$), and Q factor for various microcavities.**

|  | $\Gamma$ (meV) | $\Delta E$ (meV) | $\Omega_R$ (meV) | Q factor[a] | $\Omega/\Gamma$ |
|---|---|---|---|---|---|
| **PEAP** | 82.3 | 186 | 260 | 84 | 3.16 |
| **BAP (filled)**[b] | 99.7 | 213 (166) | 275 | 126 | 2.76 |
| **BAP (partially-filled)** | 99.7 | 207 | 190 | 83 | 1.91 |
| **Crystalline TIPS-Pc** | 128.2 | 216 | 200 | 42 | 1.56 |
| **Amorphous TIPS-Pc** | 100.9 | 105 | 160 | 38 | 1.59 |

a. Quality factor is obtained from the angle-resolved reflectance of the polariton dispersion, and is estimated at a point corresponding to 40% exciton fraction.

b. The values are from ref.[1]. Different numbers correspond to microcavities with different detuning.

## Note 5. Exciton and EP transport analysis

To analyze exciton transport, we extract the mean squared displacement MSD = $\sigma^2(t) - \sigma^2(0)$, where $\sigma^2(t)$ is the variance of the spatial distribution extracted from Gaussian fits to the expanding population profile (Figure 2d in main text, fits in Figure S6). In general, the MSD can be described by MSD = $2Dt^\alpha$, where D is diffusivity, and the exponent characterizes: subdiffusive transport ($\alpha < 1$), diffusive transport ($\alpha = 1$), superdiffusive transport ($1 < \alpha < 2$), and ballistic transport ($\alpha=2$).[3]

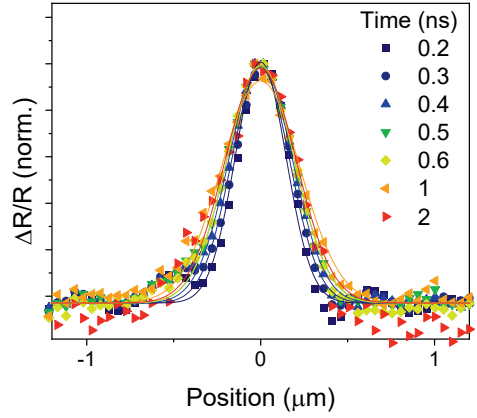

**Figure S6**. Exciton transport analysis. Bare exciton transport for PEAP with probe at ~580 nm. The pump fluence for this dataset is 5.0 µJ/cm².

For EP transport at finite $k$, $\sigma^2(t)$ can be extracted from fitting the arrival time ($t_0$) of the EP at specific positions away from the excitation spot (Figure S7 a, b), as described in our previous work.[1] To fit each arrival time across positions, we used a Gaussian function convolved with a biexponential decay.

$$\Delta R/R = A_1 e^{\left(-\frac{t}{\tau_1}+\frac{t_0+w^2}{2\tau_1^2}\right)}\left(1+erf\left(\frac{t-t_0-\frac{w^2}{\tau_1}}{\sqrt{2}w}\right)\right)$$

$$+ A_2 e^{\left(-\frac{t}{\tau_2}+\frac{t_0+w^2}{2\tau_2^2}\right)}\left(1+erf\left(\frac{t-t_0-\frac{w^2}{\tau_2}}{\sqrt{2}w}\right)\right) \quad (8)$$

where $A_1, A_2, \tau_1, \tau_2$ are amplitudes and decay lifetimes, $w$ is the IRF, and $t_0$ is the EP arrival time. An example result plotting distance vs time is shown in Figure S7c. Throughout the rest of the text, we plot MSD (rather than distance).

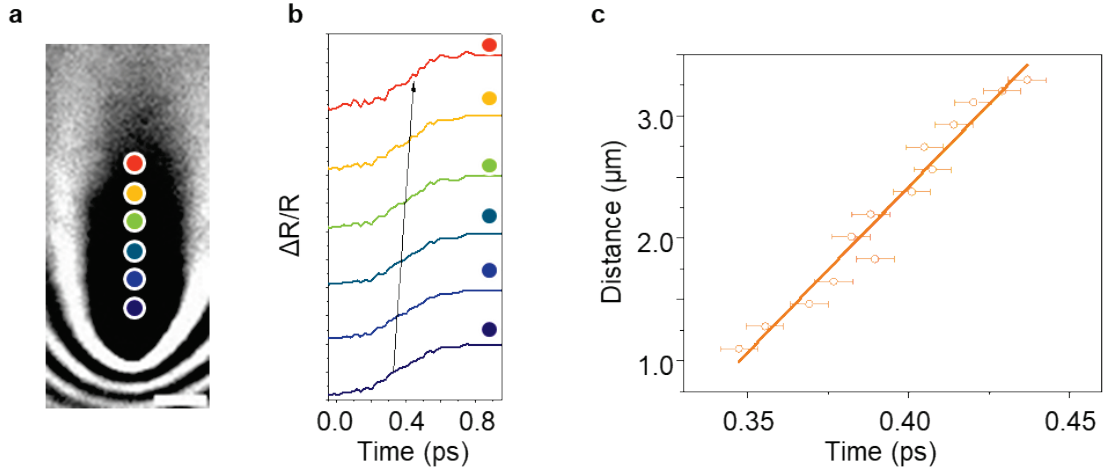

**Figure S7. Wavepacket analysis of EP transport**. **a.** EP transport imaged at $k = 10.5$ μm$^{-1}$ in a PEAP microcavity. Here the pump excitation is at the bottom of the image and propagates upward. Scale bar, 1 μm. **b**. ΔR/R profiles for each position indicated by the dots in the image (a). **c.** Distance vs. position, extracted using the wavefront arrival analysis described above. All data symbols are presented as the mean +/− one standard deviation derived from the fitting error.

**Note 6.        Additional EP transport results for PEAP microcavity**

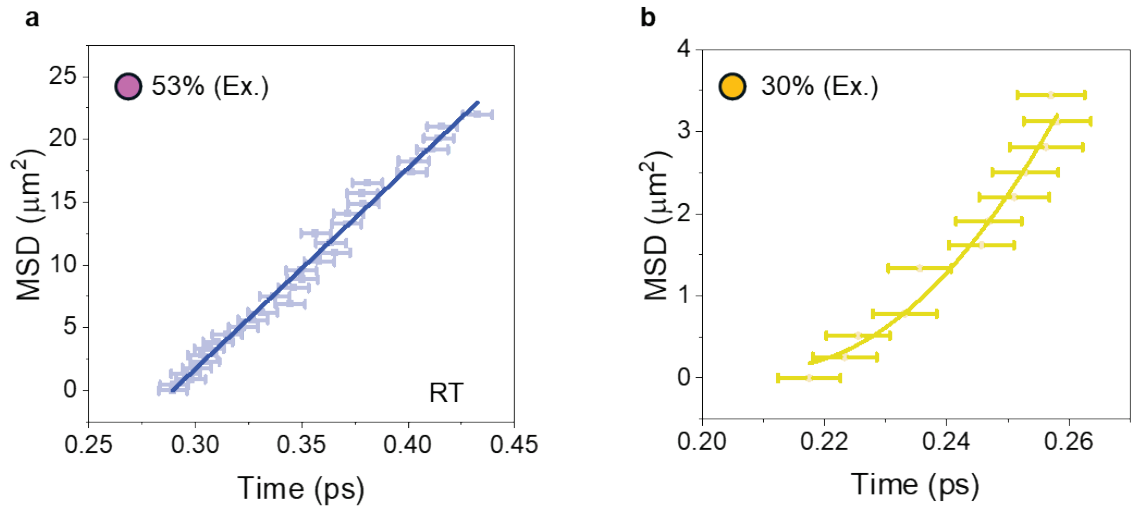

**Figure S8. Additional MSD transport results for PEAP microcavity. a-b.** MSD plots of EP transport at 2.03 eV (a, 53% exciton content) and 1.91 eV (b, 30% exciton content), extracted using wavefront analysis (Supplemental Note 5). All data symbols are presented as the mean +/− one standard deviation derived from the fitting error.

**Note 7.       Estimation of temperature-dependent exciton transfer integrals**

The zero-temperature exciton transfer integrals $J = \hbar/2m_0 a^2$ are estimated from exciton effective masses $m_0 = m_e + m_h$ (where $m_e$ is the conduction band mass and $m_h$ is the valence band mass) derived from band-structure calculations reported in the literature, and lattice constants $a$. For room-temperature $J$, we apply a temperature-dependent renormalization to the effective mass as described below. Since monomeric molecules are isolated in a polymer matrix in the amorphous TIPS-Pc samples, we assume that $J = 0$ meV.

There are multiple possible models to calculate temperature-dependent renormalization of effective masses, and the exact form of exciton-phonon interactions responsible for band renormalization in the systems studied here is not known. Here, we use a Holstein-Peierls phonon-induced renormalization model that was previously invoked to explain temperature-dependent changes to $J$ observed in angle-resolved photoemission spectra of pentacene crystals and 2D semiconductors[14,15].

$$J = J_0 \exp[-(0.5 + N_{eff})g_{eff}^2] \tag{9}$$

where $N_{eff} = [\exp(\hbar\omega_{eff}/k_B T) - 1]^{-1}$ refers to the effective phonon occupation number given by the effective phonon frequency ($\omega_{eff}$) and $g_{eff}$ is the phonon coupling strength.

The phonon coupling strength is calculated by[14]

$$g_{eff} = [\ln(1+\lambda)]^{-1} \tag{10}$$

where $\lambda = (2\pi k_B)^{-1} d\Gamma/dT$ refers to temperature-induced linewidth broadening.

**Table S2. Exciton transfer integrals $J$ in meV at zero and room temperatures.**

|  | 0 K | 295 K |
|---|---|---|
| PEAP [a] | 113 | 67.2 |
| BAP [a] | 113 | 58.6 |
| Crystalline TIPS-Pc [b] | 90.2 | 43.9 |
| Amorphous TIPS-Pc | 0 | 0 |

a. The values for $J$ (0 K) were obtained from ref.[2,16,17]. The temperature-dependent renormalization uses parameters extracted in Supplemental Note 3 and ref.[7].

b. The values for $J$ (0 K) were obtained from ref.[2,18,19]. The temperature-dependent renormalization uses parameters from pentacene crystals reported in ref.[14,15,20].

**Note 8.        Time-resolved photoluminescence of PEAP crystals**

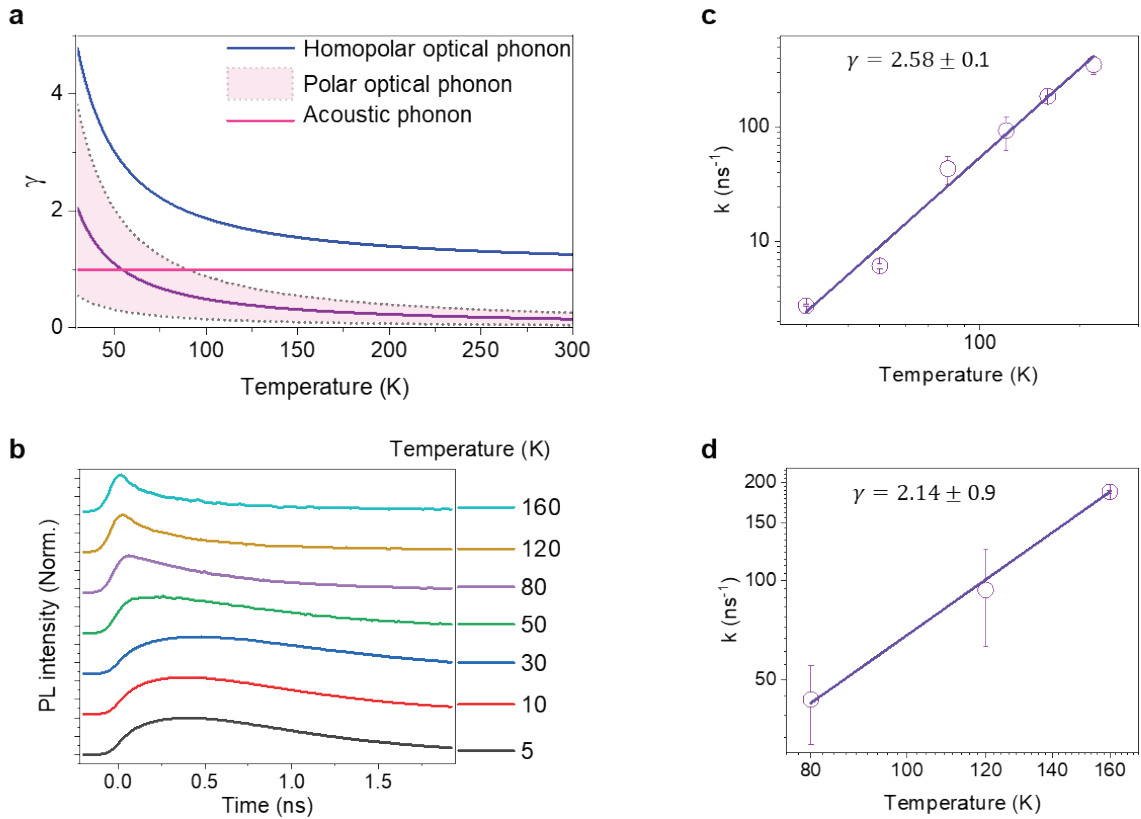

**Figure S9. Temperature dependent time-resolved photoluminescence (TRPL) of PEAP. a.** Temperature-dependent phonon interactions, described below.[21] **b.** Temperature-dependent TRPL at a pump wavelength of 515 nm and emission wavelength of 570 nm. **c-d.** Rate of rise-time extracted from panel (b) vs. temperature. The plots are fit with power law $k = T^\gamma$ where $T$ is temperature. $\gamma = 2.58$ for the range of 30-300 K (c) and $\gamma = 2.14$ for the range of 80-300 K (d). The rise time reflects primarily intraband relaxation processes to the band edge following pump excitation. All data symbols are presented as the mean +/− one standard deviation derived from the fitting error.

Following a procedure outlined in by Huang and co-workers[21], we evaluate the main phonons responsible for exciton-phonon interactions in PEAP single crystals outside microcavities using time-resolved photoluminescence (TRPL) measurements. The PL rise-time reflects intraband scattering to the band edge following pump excitation, which is primarily mediated by exciton-phonon scattering. The temperature-dependent rate of the PL rise can be fit to a power law $k = T^\gamma$, with $\gamma$ characterizing different phonon contributions as shown in Figure S9a.

By fitting the rise-time from the temperature-dependent TRPL results in Figure S9b, we extract values of $\gamma$ ranging from ~2 to 2.6 depending on the temperature range (Figures S9c,d). These values indicate that exciton-phonon interactions are dominated by homopolar phonon scattering in our PEAP crystals, as had been previously observed in other 2D perovskites.[21]

**Note 9.      Low temperature results of PEAP cavities**

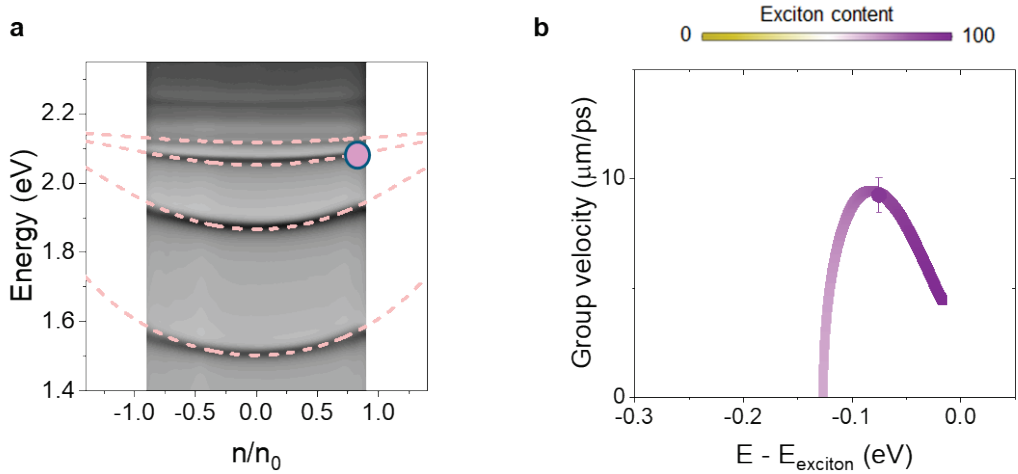

**Figure S10. Additional results for PEAP microcavity at 3.5 K. a.** Angle-resolved reflectance of polariton dispersion for a PEAP microcavity at 3.5 K. Note the lower range of momenta available due to the lower NA objective (0.9 NA) compared to RT datasets. **b.** Calculated group velocity as a function of detuning energy (E – $E_{exc}$). The measured EP velocity is overlaid with the expected group velocity.

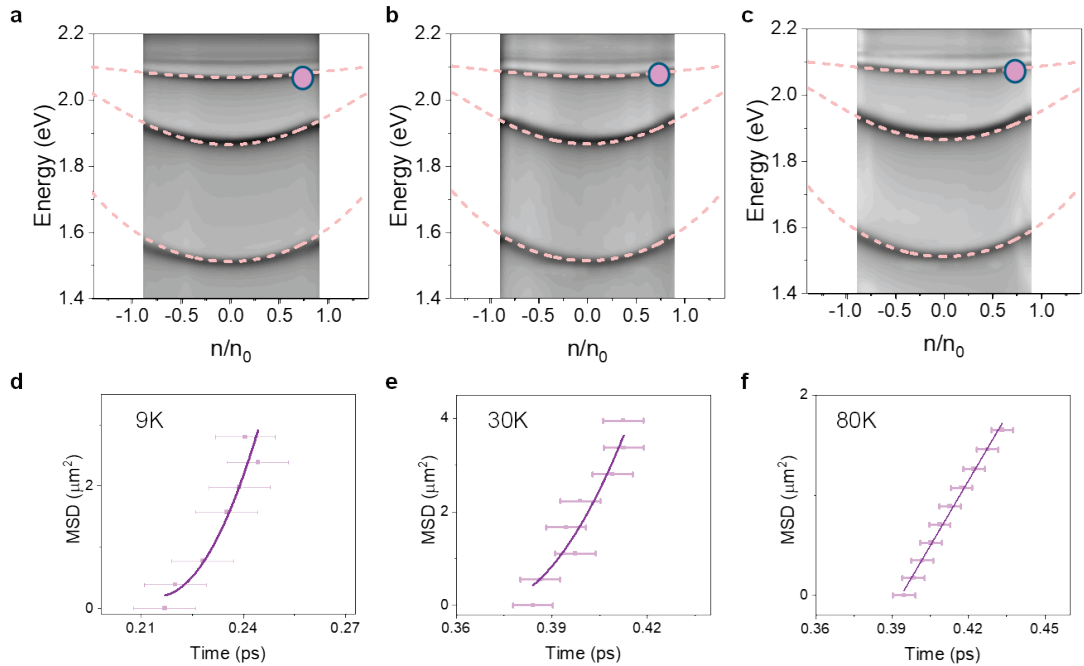

**Figure S11. Temperature dependent EP transport for PEAP microcavity with ~87% exciton content. a-c.** Angle-resolved reflectance of polariton dispersions of PEAP micocavity at 9 K (a), 30 K (b), and 80 K (c). Dashed lines are coupled oscillator model fits used to extract the Hopfield coefficients for each temperature. **d-f.** MSD plots of EP transport at 9 K (d), 30 K (e), and 80 K (f). All data symbols are presented as the mean +/− one standard deviation derived from the fitting error.

# Note 10. Amorphous and Crystalline TIPS-PC films

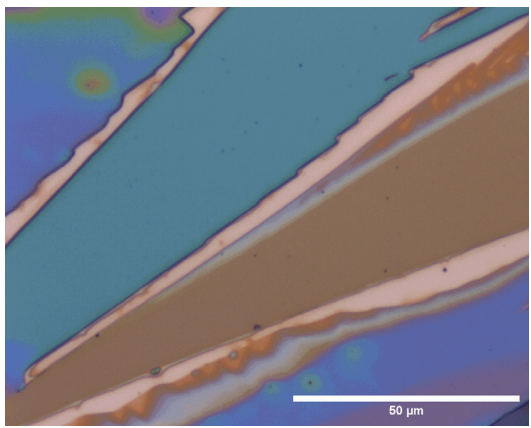

**Figure S12.** Optical image of crystalline TIPS-Pc domains. Scale bar, 50 µm.

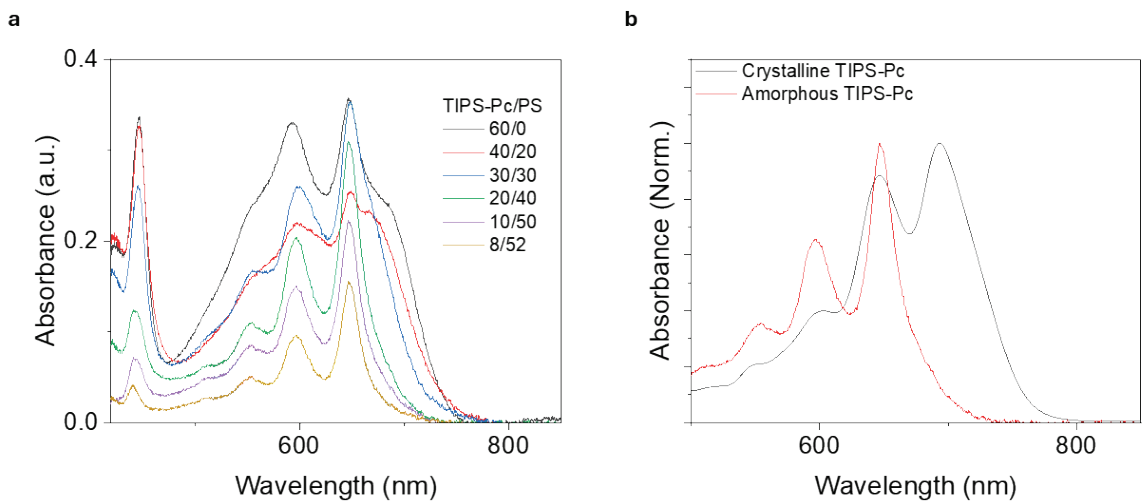

**Figure S13. Absorbance of TIPS-Pc samples. a.** Concentration dependent absorption spectra of amorphous TIPS-Pc films in polystyrene using various concentration ratios (reported in the legend as mg/mg in 1 ml toluene solutions). **b.** Normalized absorbance of crystalline and amorphous TIPS-Pc films.

As shown in Figure S13a, amorphous TIPS-Pc thin films show signs of aggregation at concentrations exceeding 20 mg/ml. Thus, all measurements are performed at concentrations of 20 mg/ml for the EP measurements in the text to ensure samples representative of isolated monomeric molecules in a polymer matrix. In contrast, the absorption spectrum of crystalline TIPS-Pc (Figure S13b) shows character consistent with mixing of Frenkel and charge transfer excitons typical for crystalline domains of this material.[22]

## Note 11. EP transport for crystalline TIPS-Pc microcavity at RT

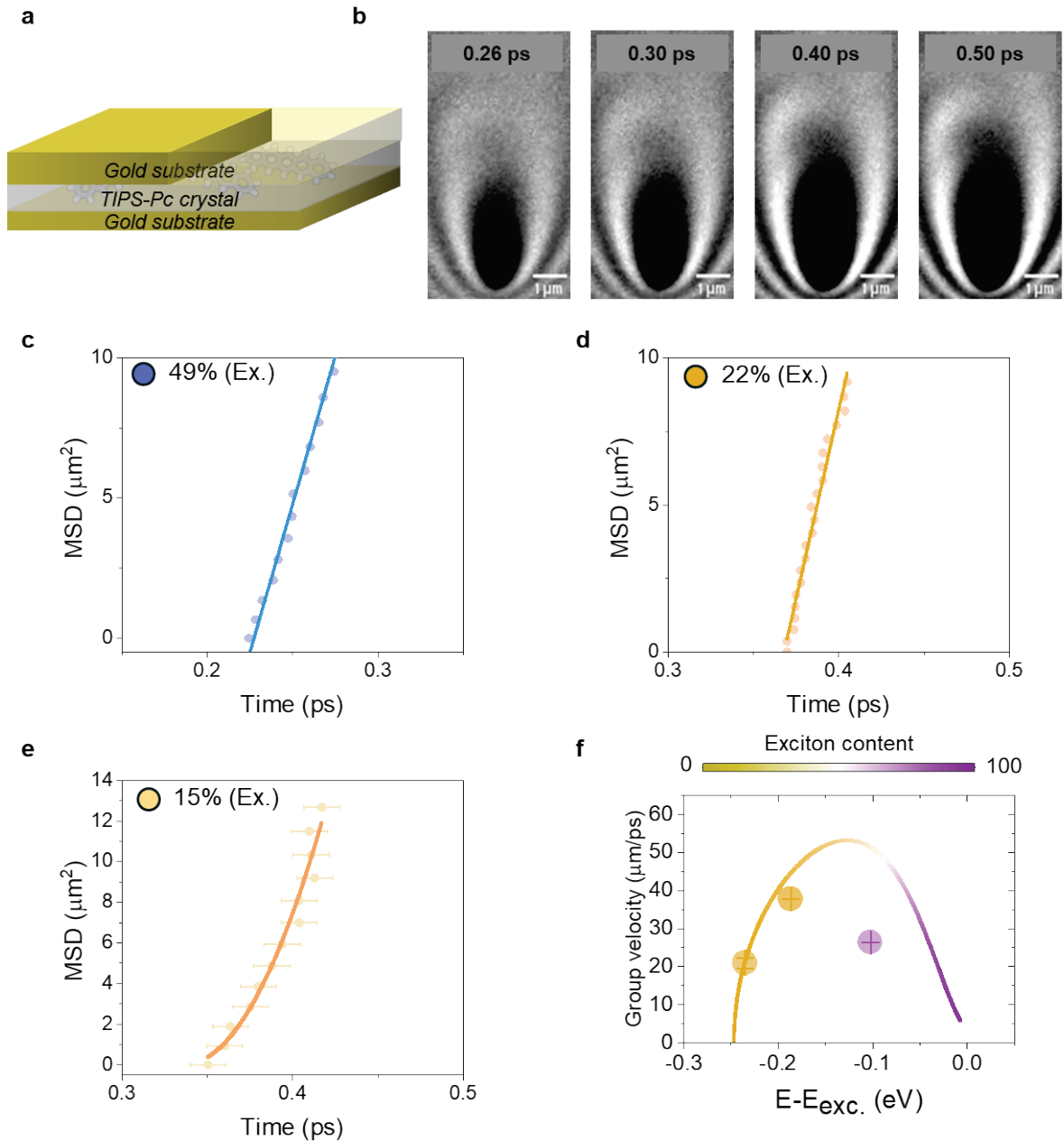

**Figure S14. MUPI results for crystalline TIPS-Pc microcavity at room temperature. a.** Schematic structure for crystalline TIPS-Pc microcavity. **b.** EP transport imaged at $k = 3.0$ μm$^{-1}$, corresponding to EP with 15% excitonic content. **c-e**. MSD plots of EP transports at 1.65 eV (c, 49% exciton content), 1.59 eV (d, 22% exciton content) and 1.53 eV (e, 15% exciton content), extracted using wavefront analysis (Supplemental Note 5). **f.** Calculated group velocity as a function of detuning energy (E – E$_{exc}$). The measured EP velocities are overlaid with the expected group velocity. The measured group velocities are probed at the points labeled with circle in Figure 3a in main text. All data symbols are presented as the mean +/− one standard deviation derived from the fitting error. All scale bars, 1 μm.

## Note 12. EP transport for amorphous TIPS-Pc microcavity at RT

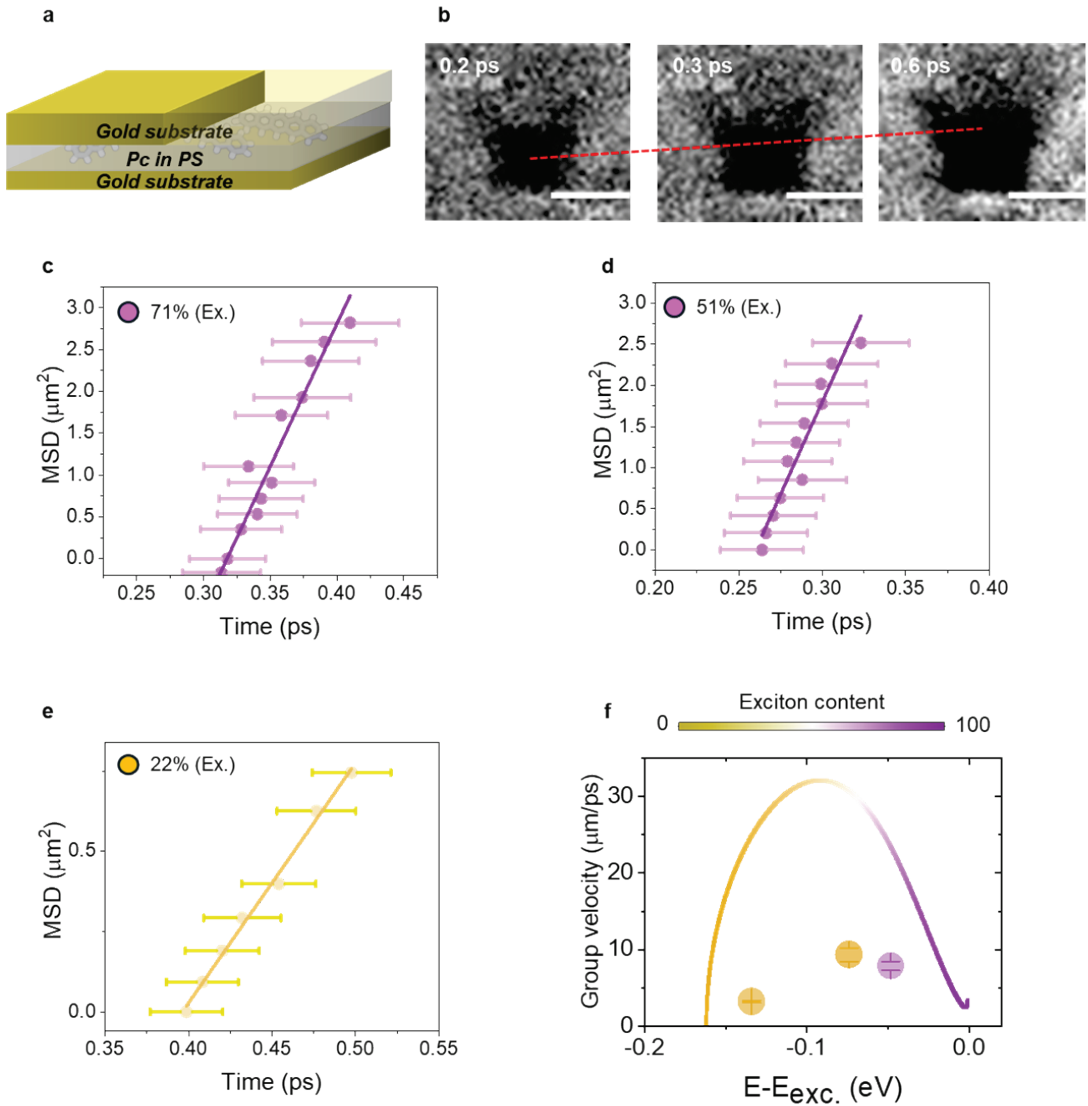

**Figure S15. MUPI results for amorphous TIPS-Pc microcavity at room temperarure. a.** Schematic structure for amorphous TIPS-Pc microcavity. **b.** EP transport imaged at $k = 4.8\ \mu m^{-1}$, corresponding to EPs with 51% excitonic content. **c-e**. MSD plots of EP transport at 1.85 eV (c, 71% exciton content), 1.82 eV (d, 51% exciton content) and 1.75 eV (e, 22% exciton content), extracted using wavefront analysis (Supplemental Note 5). **f.** Calculated group velocity as a function of detuning energy (E – $E_{exc}$). The measured EP velocities are overlaid with the expected group velocity. The measured group velocities are probed at the points labeled with circle in Figure 3b in the main text. All data symbols are presented as the mean +/− one standard deviation derived from the fitting error. All scale bars, 1 μm.

Note 13.        EP transport for crystalline TIPS-Pc microcavity at 5 K.

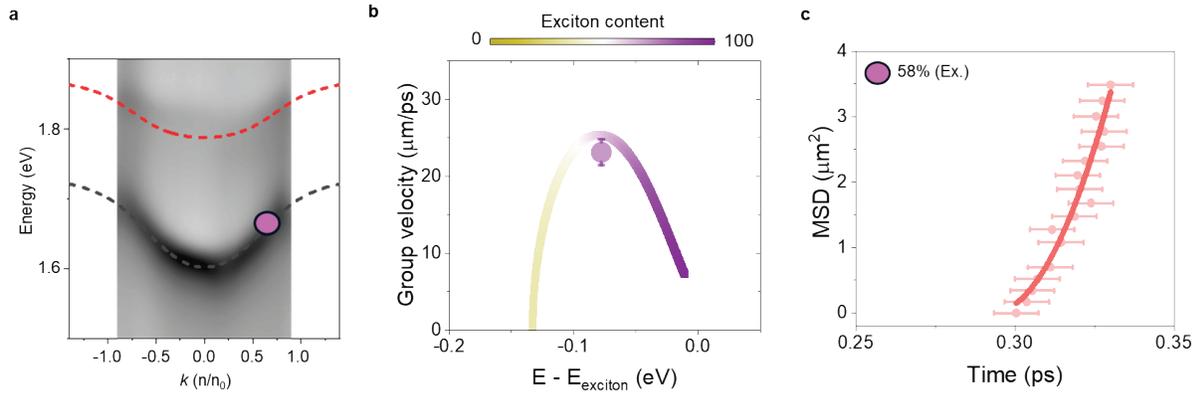

**Figure S16. Cryogenic temperature results for crystalline TIPS-Pc microcavity. a.** Angle-resolved reflectance of polariton dispersion for a crystalline TIPS-Pc microcavity at 5 K. Dashed lines are coupled oscillator model fits. Middle polariton features are beyond the NA of the in-cryostat objective. **b.** Calculated group velocity as a function of detuning energy ($E - E_{exc}$). The measured EP velocities are overlaid with the expected group velocity. The measured group velocities are probed at the point labeled with a circle in Figure S16a. **c.** MSD of EP transport probed at 1.65 eV, corresponding to EP with 58% exciton content, extracted using wavefront analysis (Supplemental Note 5). All data symbols are presented as the mean +/− one standard deviation derived from the fitting error.

Note 14.        EP transport for amorphous TIPS-Pc microcavity at 5 K

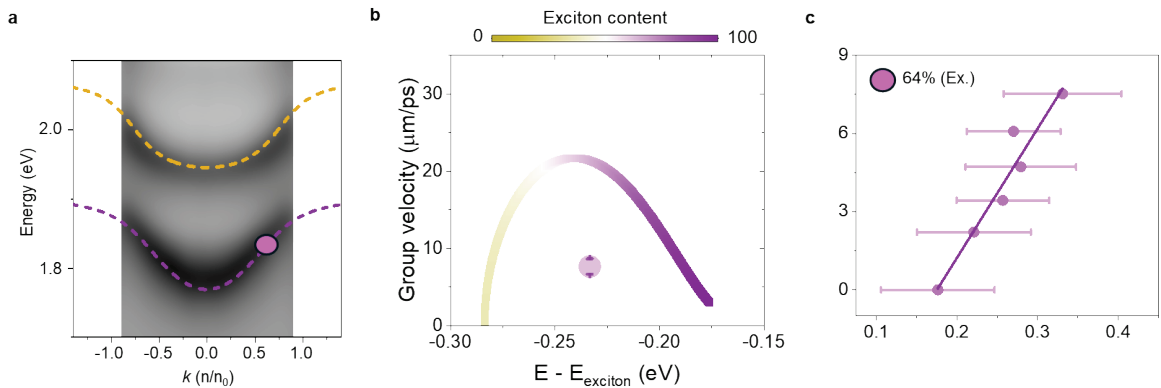

**Figure S17. Cryogenic temperature results for amorphous TIPS-Pc microcavity. a.** Angle-resolved reflectance of polariton dispersion for amorphous TIPS-Pc microcavity at 5 K. Dashed lines are coupled oscillator model fits. Middle polariton features are beyond the NA of the in-cryostat objective. **b.** Calculated group velocity as a function of detuning energy ($E - E_{exc}$). The measured EP velocities are overlaid with the expected group velocity. The measured group velocities are probed at the point labeled with the circle in Figure S17a. **c.** MSD plots of EP transport probed at 1.84 eV, corresponding to EP with 64% exciton content, extracted using wavefront analysis (Supplemental Note 5). All data symbols are presented as the mean +/− one standard deviation derived from the fitting error.

**Note 15.    Low temperature absorbance for crystalline and amorphous TIPS-Pc**

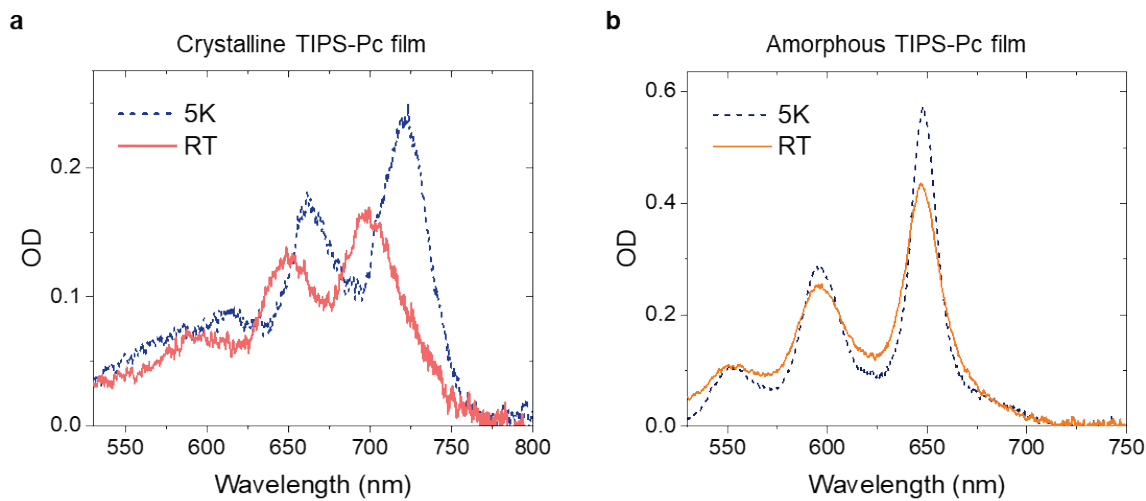

**Figure S18. Temperature dependent absorption spectra. a.** Absorbance of crystalline TIPS-Pc at 5 K (dashed line) and RT (solid line). **b.** Absorbance of amorphous TIPS-Pc at 5K (dashed line) and RT (solid line).

## Note 16. Correlating EP velocity renormalization to various parameters

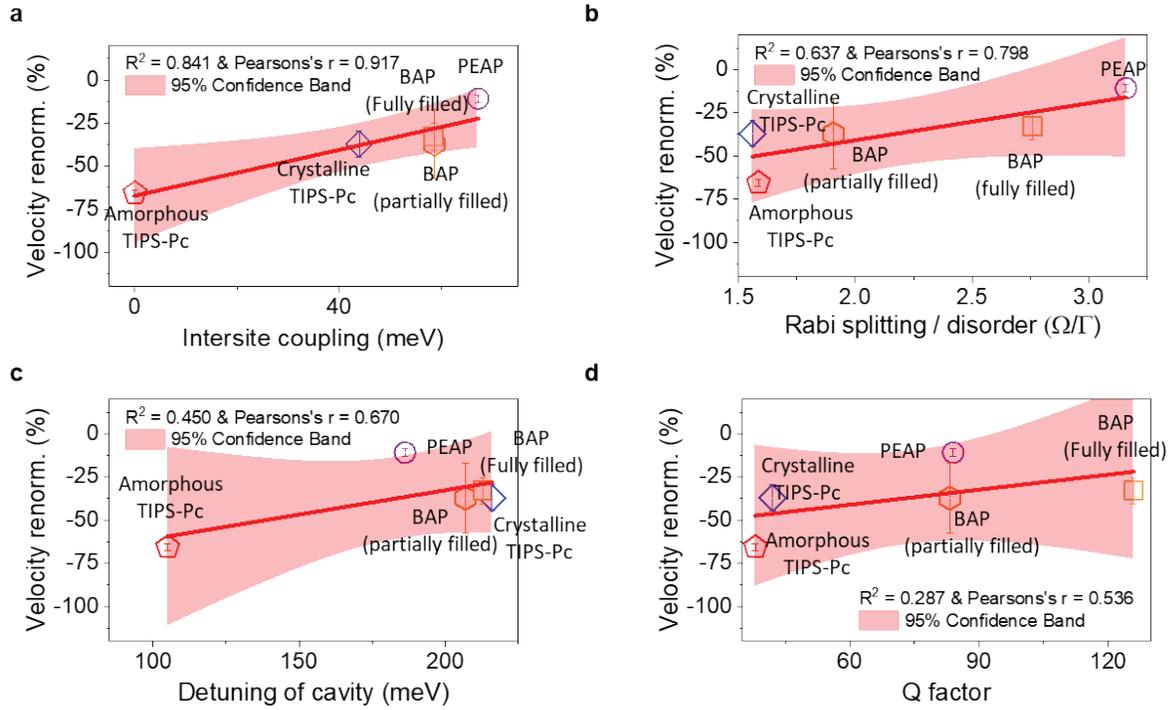

**Figure S19**. **Correlation of EP velocity renormalization with various parameters. All data correspond to EPs with 40% exciton fraction at room temperature**. **a.** EP velocity renormalization as a function of intersite coupling. **b.** EP velocity renormalization as a function of Rabi splitting/disorder. **c**. EP velocity as a function of cavity-exciton detuning. **d.** EP velocity renormalization as a function of Q factor. The symbols correspond to respective microcavities: PEAP (purple circle), BAP (yellow square for fully-filled and yellow hexagon for partially-filled), crystalline TIPS-Pc (blue diamond), amorphous TIPS-Pc (red pentagon) cavities. The red shaded region indicates the 95% confidence band. All data symbols are presented as the mean +/− one standard deviation derived from errors in the linear fit of renormalization vs exciton content.

Of all parameters explored in Figure S19, we find that EP renormalization correlates best with intersite coupling (Figure S19a). While other parameters (S19c-d) show weak correlation, we note that the quantity $\Omega_R/\Gamma$ (Figure S19b) also displays relatively good correlation. To rule out that this parameter governs EP renormalization in the microcavities explored here, we perform an additional control experiment that varies $\Omega_R$ while keeping all other parameter constant. We fabricate a partially-filled BAP microcavity using a thin BAP structure encapsulated in polymer spacers (Figure S20). This structure remains in the strong coupling regime, but reduces $\Omega_R$ from 275 meV in a fully-filled cavity to 190 meV in the partially-filled cavity. We observe very similar EP transport as in the fully-filled cavity (Figure S20), suggesting that $\Omega_R$ for such layered materials in the strong-coupling regime is not a good predictor of EP renormalization.

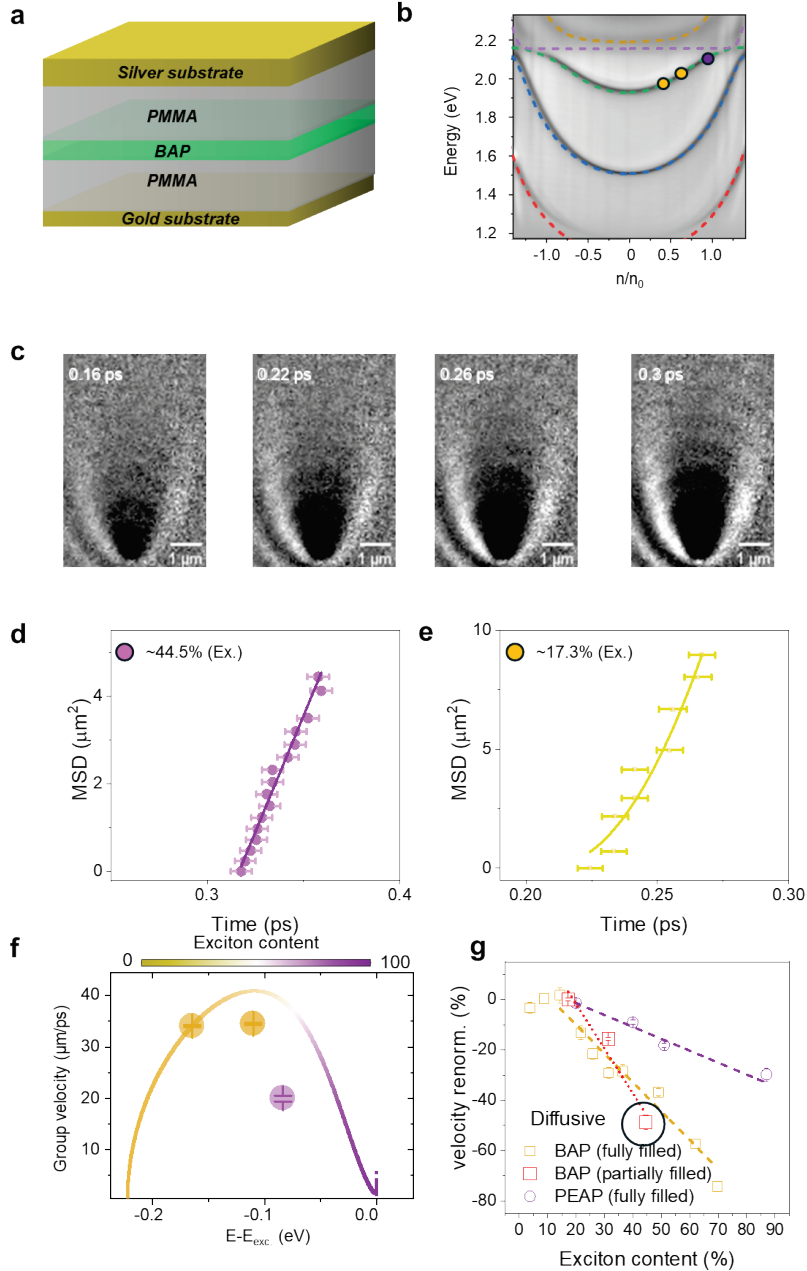

**Figure S20. MUPI results for partially filled BAP microcavity. a.** Schematic structure for partially filled BAP microcavity[9]. **b.** Angle-resolved reflectance of polariton dispersion for a partially filled BAP microcavity. A coupled oscillator fit (Supplemental Note 4) yields $\Omega_R$= 190 meV. **c.** EP transport imaged at 1.95 eV, corresponding to EPs with 25% excitonic fraction. **d-e**. MSD plots of EP transport probed at 2.05 eV (d, 44.5% exciton content) and 1.95 eV (e, 17.3% exciton content), extracted using wavefront analysis (Supplemental Note 5). **f.** Calculated group velocity as a function of detuning energy ($E - E_{exc}$). The measured EP velocities are overlaid with the expected group velocity. The measured group velocities are probed at different points labeled with circle in Figure S20b. **g.** Velocity renormalization plots overlapping BAP (square for fully filled and partially filled), PEAP (circle). All data symbols are presented as the mean +/− one standard deviation derived from the fitting error. All scale bars, 1 μm.